\documentclass[12pt]{article}

\usepackage{scicite}
\usepackage{graphicx}
\usepackage{times}
\usepackage{amssymb}

\newenvironment{sciabstract}{%
\begin{quote} \bf}
{\end{quote}}

\title{Comet Fading Begins Beyond Saturn}

\author{
Nathan A. Kaib$^{1\ast}$\\
\normalsize{$^{1}$Department of Physics \& Astronomy,}\\
\normalsize{440 W. Brooks St, Norman, OK 73019, USA}\\
\\
\normalsize{$^\ast$To whom correspondence should be addressed; E-mail:  nathan.kaib@ou.edu.}
}

\date{}

\begin{document} 


\baselineskip24pt


\maketitle 


\begin{sciabstract}
The discovery probability of long-period comets (LPCs) passing near the Sun is highest during their first passage and then declines, or fades, during subsequent return passages. Comet fading is largely attributed to devolatilization and fragmentation via thermal processing within 2--3 au of the Sun (1 au being the Earth-Sun distance). Here our numerical simulations show that comet observing campaigns miss vast numbers of LPCs making returning passages through the Saturn region (near 10 au) because these comets fade during prior, even more distant passages exterior to Saturn and thus elude detection. Consequently, comet properties significantly evolve at solar distances much larger than previously considered, and this offers new insights into the physical and dynamical properties of LPCs, both near and far from Earth.
\end{sciabstract}

{\bf Teaser:} Long-period comets are discovered to fade as they make repeated passages through the region beyond Saturn.

\section*{Introduction}
Because of the gravitational barriers of Jupiter and Saturn, LPCs must attain semimajor axes ($a$) in excess of $\sim$20,000 au before perturbations from the Galactic tide and passing stars can torque their perihelia ($q$, or distance of closest approach to the Sun) from the outer solar system to near the Earth \cite{kaibquinn09,hills81}. During their first perihelion passage through the inner solar system, energy kicks from the gas giants will either eject these ``new'' LPCs on hyperbolic orbits or shrink their semimajor axes to much smaller values, allowing them to make one or more return passages near the Sun and Earth. Owing to their shorter orbital periods and potential for multiple near-Sun passages, we might then expect ``returning'' LPCs to be much more commonly observed than new LPCs making their first passage through the inner solar system \cite{wietre99}. However, this is not the case \cite{oort50}; LPCs on new orbits (for which the formal classification criterion is $a>10^4$ au) comprise 45\% of all discovered LPCs with well-determined orbits \cite{mars08}. The reason this number is disproportionately large relative to dynamical expectations is because higher near-surface volatile content of new LPCs enhances cometary activity, maximizing their discovery probability \cite{oortschmidt51}. In contrast, thermal processing during prior perihelion passages diminishes the brightness of returning LPCs as they devolatilize and fragment, or ``fade,'' during each orbital revolution \cite{weiss79,nes07,vok19}.

Comets with perihelia in the inner few au of the solar system dominate the historical catalog of known LPCs, as their near-Sun activity and proximity to Earth enhances discovery probability \cite{mars08}. A large fraction of these comets cannot be physically or dynamically pristine, as they are subjected to near-Sun fading and dynamical processing via pre-discovery encounters with Jupiter and Saturn \cite{kaibquinn09}. Thus, a sample of LPCs with perihelia well beyond Jupiter has long been desired, as their orbits are less altered by planetary perturbations and their surfaces less subjected to thermal processing.

\section*{Results}
The most distant known LPC perihelion (comet C/2003 A2) is now 11.42 au, and the JPL small-body database contains 25 LPCs with perihelia between 8 and 12 au. The distribution of their original semimajor axes (their semimajor axis values before their current passage through the planetary region) is shown in Fig. 1. With 76\% of these comets on semimajor axes over $10^4$ au, these distant LPCs' semimajor axes are even more heavily skewed toward dynamically new orbits than the historical LPC catalog. 

The overabundance of $a>10^4$ au comets at near-Saturn perihelia is surprising for two reasons. First, if comet fading only occurs during perihelion passages through the inner solar system, we should expect returning LPCs to comprise a much greater fraction of comet discoveries at these perihelion distances. Second, the gravitational perturbations of Jupiter and Saturn are less effective at ejecting or altering comet semimajor axes with such large perihelia, and we should expect still more discoveries of LPCs on smaller ($a<10^4$ au) orbits \cite{hills81,siltre16,vok19}. 

\begin{figure}[htbp]
\centering
\includegraphics[scale=0.8]{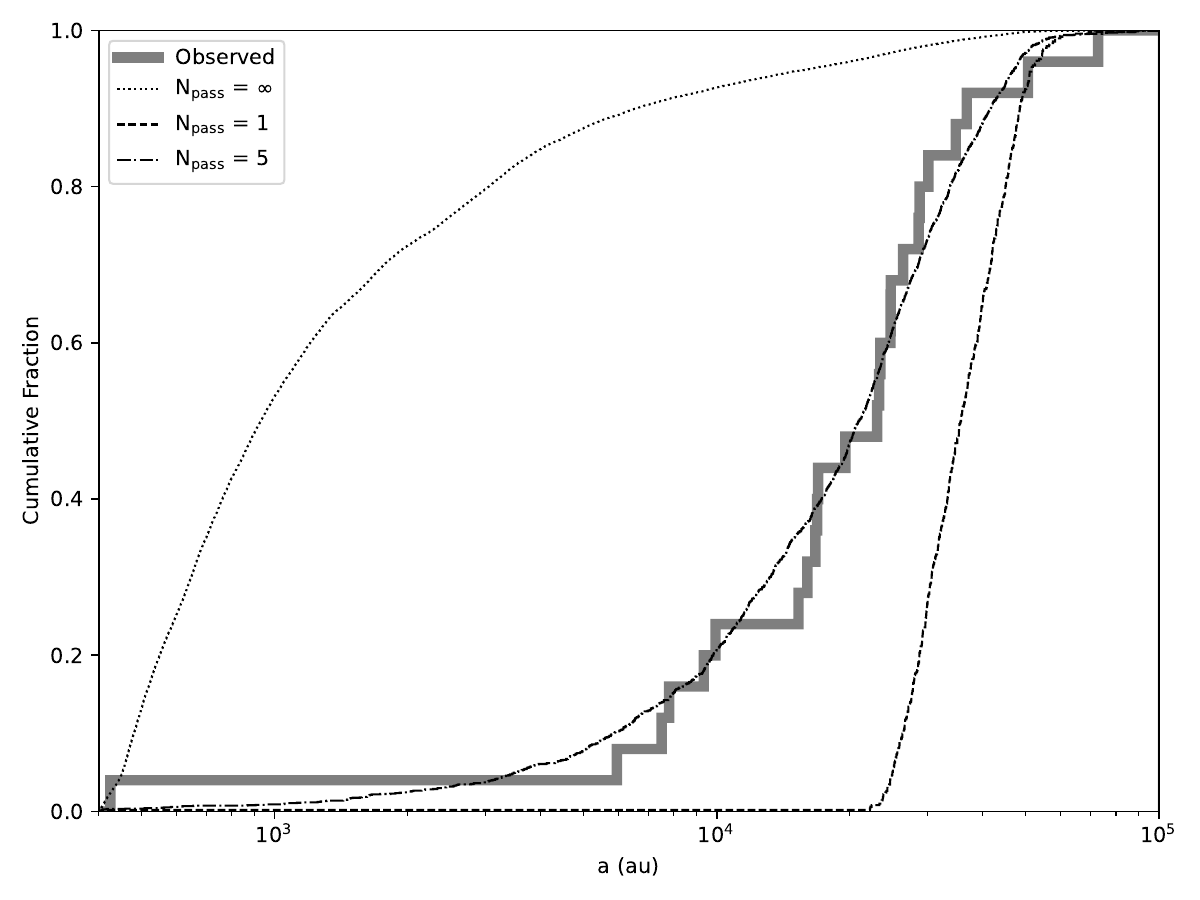}
\caption{{\bf Observed and simulated original semimajor axes of distant long-period comets.} Cumulative distribution of original semimajor axes for LPCs with $8 <q <12$ au and $a>400$ au. Solid line marks observed LPCs. Non-solid lines are simulated LPCs assuming no fading ({\it dotted}), fading after 1 passage inside 20 au ({\it dashed}), and fading after 5 passages inside 20 au ({\it dash-dotted}).}
\end{figure}

To illustrate the unexpected nature of these LPC observations, we run an N-body simulation modeling LPC production as the local Galaxy perturbs distant Oort cloud bodies into planet-crossing orbits (see supporting online material for a detailed description). From our simulation, we build a synthetic LPC sample by assuming that every simulated body will be detected during perihelion passages below 12 au. The dotted line in Fig. 1 shows the semimajor axis distribution of LPCs with $8<q<12$ au predicted by our simulation. The simulated distribution is radically different from the observed. 96\% of simulated comets pass through perihelion on semimajor axes below $10^4$ au, compared to 24\% of observed. Analysis via the binomial probability distribution indicates a less than 1 in $10^{20}$ chance that the heavier skew toward $a>10^4$ au amongst observed LPCs with $8<q<12$ au (compared to simulations) is a result of the small observed sample size of distant LPCs. The match of simulations to observations can be dramatically improved if we assume that LPC brightness rapidly fades as comets make successive perihelion passages within 20 au of the Sun. Assessing fitness via a Kolmogorov-Smirnov test, our best match to observed data ($p$-value of 0.85) is achieved when simulated LPCs fade below detectability after just 5 passages inside 20 au. Without fading, the typical detected LPC spends $\sim$10 Myrs with $q<20$ au and undergoes over 50 passages inside 20 au during that timespan.

Rapid LPC fading should require that the perihelia of most observed LPCs are quickly moving Sunward. (Otherwise, the comets would have already faded below detectability during past perihelion passages.) Outside of rare, powerful stellar passages \cite{kaib11b}, the rate of LPC perihelion change is usually dictated by the Galactic tide, and this is proportional to $-\sin 2\omega_G$, where $\omega_G$ is the orbital argument of perihelion measured relative to the Galactic midplane \cite{heitre86}. Thus, if $\sin 2\omega_G$ is positive, then the Milky Way tide drives perihelion toward the Sun. If $\sin 2\omega_G < 0$, perihelion is pulled away from the Sun.

\begin{figure}
\centering
\includegraphics[scale=0.5]{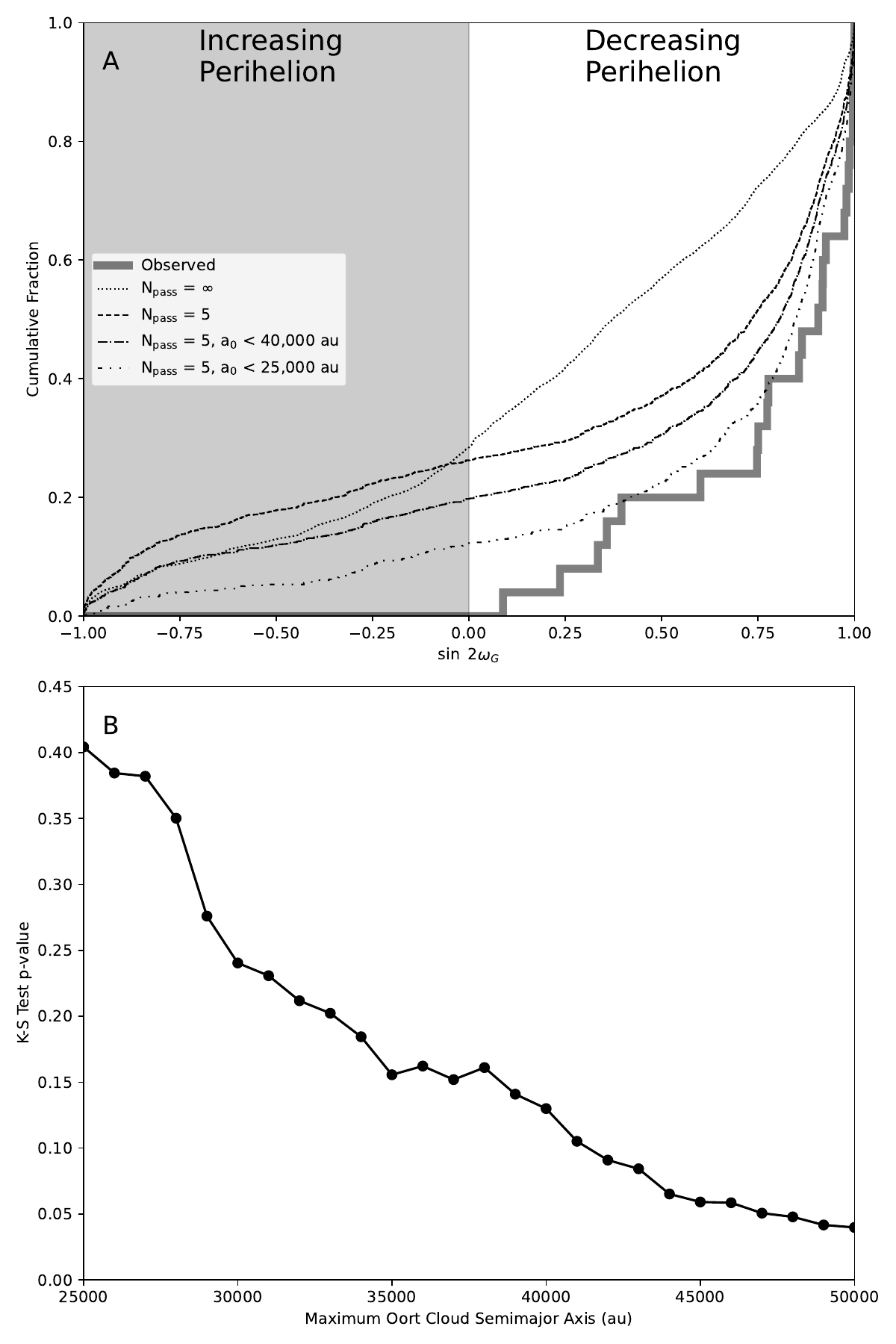}
\caption{{\bf Observed and simulated arguments of perihelion of distant long-period comets.} {\bf A:} Cumulative distribution of sine of twice the argument of perihelion (with respect to the Milky Way midplane) for LPCs with $8 <q <12$ au and $a>400$ au. Solid line marks observed LPCs. Non-solid lines are simulated LPCs assuming no fading ({\it dotted}), fading after 5 passages inside 20 au ({\it dashed}), fading after 5 passages inside 20 au but excluding comets with initial semimajor axes beyond 40,000 au ({\it dash-dotted}), and fading after 5 passages inside 20 au but excluding comets with initial semimajor axes beyond 25,000 au ({\it dash-dot-dotted}). {\bf B:} The $p$-value of a K-S test comparing $\sin 2\omega_G$ of observed and simulated LPCs (with $8 <q <12$ au and $a>400$ au) is plotted against the initial maximum semimajor axis of our simulated Oort cloud from which simulated LPCs are generated.}
\end{figure}

In Fig. 2, we plot the $\sin 2\omega_G$ distribution for all known LPCs with $8<q<12$ au. We see that the distribution shows extreme clustering toward +1, without a single negative value. Although observational survey biases are undoubtedly present, the extreme skew of the observed $\sin 2\omega_G$ distribution is highly consistent with rapid LPC fading beyond the Saturn region, since it implies that final pre-discovery perihelion passages of known LPCs were generally further from the Sun. In other words, LPCs with $\sin 2\omega_G > 0$ are much more likely to be discovered than comets with $\sin 2\omega_G < 0$ because the former LPC group are typically making closer and closer successive perihelion passages, and new near-surface species are being sublimated for the first time on each perihelion passage. This is not the case for the $\sin 2\omega_G < 0$ LPC group, whose arguments of perihelion imply that they generally have previously undergone closer perihelion passages.

The evidence for rapid LPC fading is further supported when we examine our simulated LPC orbits with $8<q<12$ au in Fig. 2. Without LPC fading, we predict a broad $\sin 2\omega_G$ distribution, with negative values for 28.4\% of discovered LPCs. Assuming this fraction of simulated negative values, there is only a 1 in $\sim$4000 chance that our 25 observed LPCs should all have positive $\sin 2\omega_G$, according to a binomial probability distribution. If we employ Fig. 1's best fit fading law, the augmented distribution clusters much more closely toward +1, and the negative fraction falls slightly to 26.2\%. This is still not as extreme as the observed distribution, and we might expect the match to observations to markedly improve if we adopt even more extreme comet fading. However, this is not the case. If we assume LPCs fade after just 1 passage inside 20 au, the negative fraction of $\sin 2\omega_G$ only falls to 25.2\%. The reason for this is that the perihelia of large ($a\sim$ 50,000 au) semimajor axis LPCs fluctuate rapidly over one orbital revolution, reaching a minimum value and then rebounding away from the Sun by the time they actually make their first $q<20$ au perihelion passage.

Thus, Fig. 2's simulated distribution of $\sin 2\omega_G$ should also be sensitive to the radial extent of our simulated Oort cloud, which is initially populated with a power-law radial number density profile \cite{vok19} from 20,000 $<a<$ 50,000 au. If we instead only consider simulated LPCs with initial semimajor axes below 40,000 au, Fig. 2 now shows that only 21.7\% of the $\sin 2\omega_G$ distribution is below zero (assuming the preferred fading law of Fig. 1), and clustering toward +1 is even further enhanced. A K-S test cannot reject this new $\sin 2\omega_G$ distribution ($p=0.13$), and we also find that this semimajor axis augmentation does not notably degrade Fig. 1's match to observations.

In Fig. 2B, we plot the $p$-value yielded by a K-S comparison between simulated and observed $\sin 2\omega_G$ values as a function of the maximum semimajor axis of our simulated Oort cloud. (This analysis continues to assume that LPCs fade after 5 passages inside 20 au.) Here we see that the match of simulated $\sin 2\omega_G$ values to observed ones steadily improves as we decrease the maximum semimajor axis of the Oort cloud. If we restrict our analysis to LPCs with initial semimajor axes below 25,000 au, the negative fraction of simulated $\sin 2\omega_G$ values is only 12.5\% and the $p$-value returned by a K-S test is 0.43. The actual radial structure of the Oort cloud is uncertain and tied to the Sun's birth environment and dynamical history, and many Oort cloud formation models predict steeper profiles than ours with ample bodies interior to $a<$ 20,000 au \cite{dqt87,fern97,dones04,bras06,kaib11b}. It is likely that our match to observed $\sin 2\omega_G$ values could be further improved if we included an Oort cloud population below $a<$ 20,000 au. While likely unphysical, the decision to truncate our cloud's inner edge at 20,000 au is conservative because the $a<$ 20,000 au population is particularly poorly constrained, and its presence will exacerbate the need for LPC fading, as outlined in our supporting online text, potentially making the necessity of fading a foregone conclusion.

\begin{figure}
\centering
\includegraphics[scale=0.8]{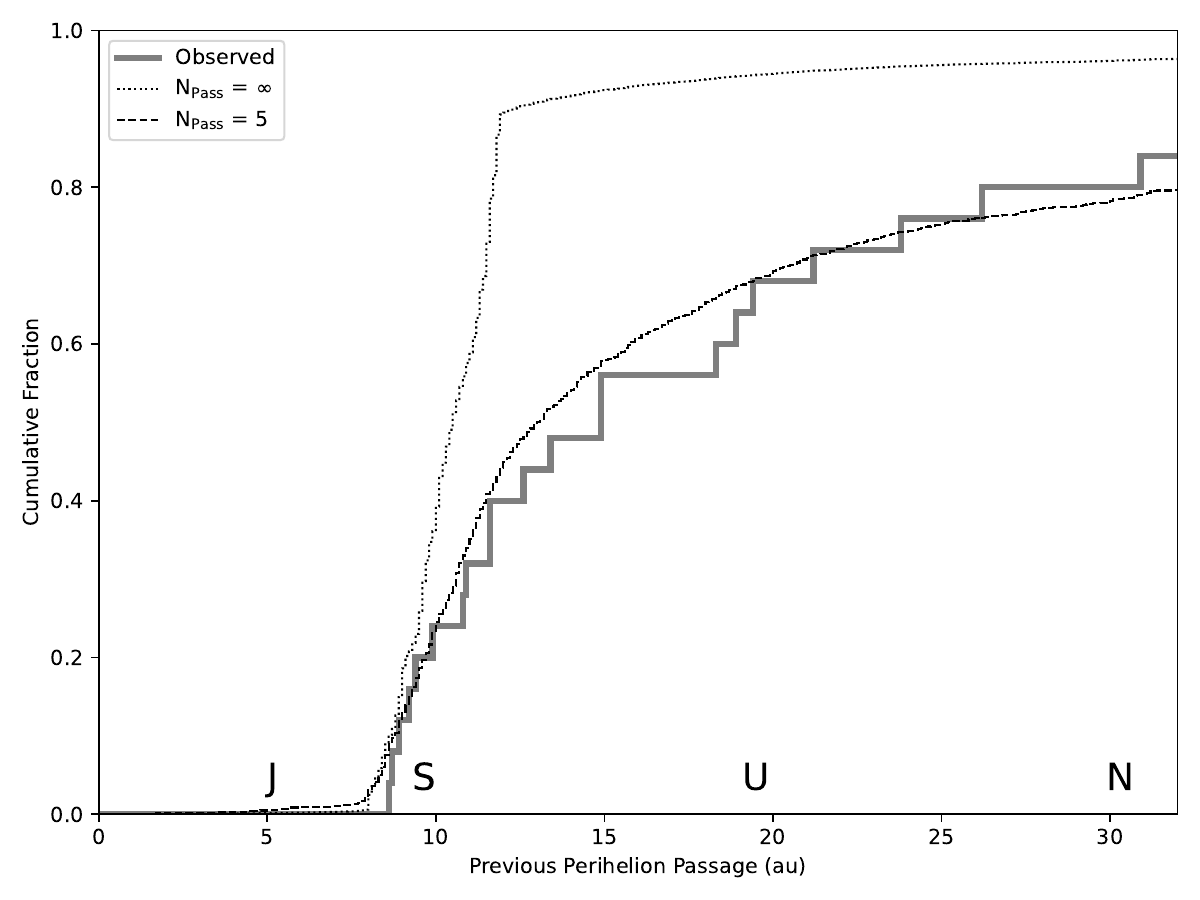}
\caption{{\bf Estimated previous perihelion passage distances of observed and simulated distant long-period comets.} Cumulative distribution of the predicted previous perihelion passage distances for LPCs with $8 <q <12$ au and $a>400$ au. Solid line marks observed LPCs. Non-solid lines are simulated distributions assuming no fading ({\it dotted}) and fading after 5 passages inside 20 au ({\it dashed}). }
\end{figure}

Through back-integrating their orbits, we can also study the distribution of observed LPCs' final, pre-discovery perihelion passage distances. Since the exact stellar encounter history of the solar system is uncertain, this method only considers the Galactic tide, yet it provides accurate estimates as long as the solar system has not undergone an exceptionally powerful stellar encounter in the past few Myrs \cite{weiss96,bail18}. These back-integration results are shown in Fig. 3. Consistent with rapid LPC fading beyond Saturn, previous perihelion passages of the observed LPCs are generally far from the Sun. 19 of our 25 known LPCs had prior perihelion passages further than $\sim$11 au from the Sun, and the 6 that did not happen to be the only observed LPCs with $a<10^4$ au. According to Fig. 1, our best-fit fading law shows observed LPCs with $a<10^4$ au are overwhelmingly likely to be returning (but not yet faded) LPCs rather than than new LPCs on their first perihelion passage. 
 
We next perform the same back-integration of our simulated LPCs with $8<q<12$ au, with the results overplotted in Fig. 3. Once again, we see that the simulated distribution is wildly different from the observed one unless comet fading is implemented. When we again assume LPCs fade after 5 passages inside 20 au, the simulated distribution resembles the form of the observed one, and a K-S test cannot reject the null hypothesis ($p=0.93$).

\begin{figure}
\centering
\includegraphics[scale=0.4]{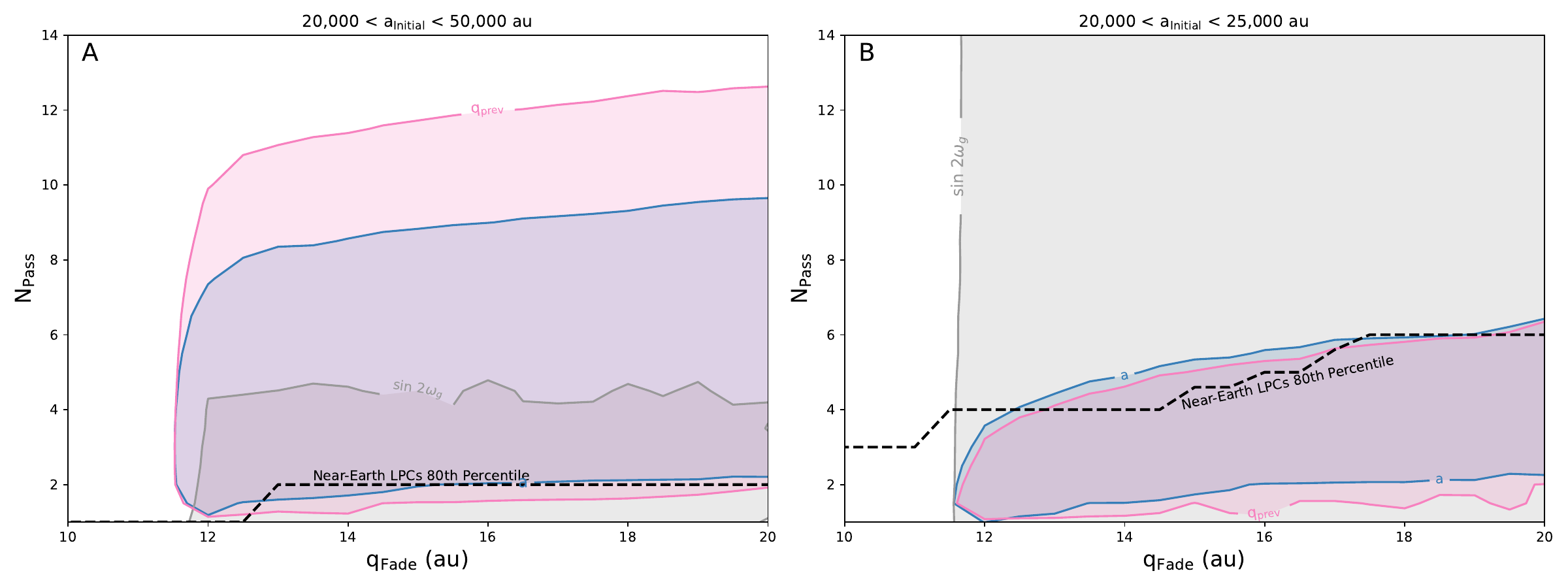}
\caption{{\bf Distant long-period comet fading parameter space permitted by known comets and simulated comet production.} Number of perihelion passages before comets fade are plotted against the perihelion distance inside which fading operates for LPCs with $8<q<12$ au. Filled contours mark area outside of which a K-S test rejects (with over 2$\sigma$ confidence) the null hypothesis that simulated and observed LPCs have the same underlying distribution. Contours are calculated for LPC original semimajor axis, $\sin 2\omega_G$, and estimated previous perihelion passage. For simulated new ($a>10^4$ au) near-Earth ($q<4$ au) LPCs, we also plot the 80th percentile of the number of passages made inside a given perihelion prior to inner solar system entry ({\it dashed line}). Panel {\bf A} considers all simulated comets and Panel {\bf B} excludes those initialized on semimajor axes beyond 25,000 au.}
\end{figure}

Our first three figures demonstrate how effectively one particular fading law can replicate distant LPC observations, but the perihelion distance at which fading begins as well as the number of perihelion passages before comets fade are both physically unconstrained parameters. Thus, for all combinations of these two parameters, we can repeat the analyses of Figs. 1--3, using a K-S test to compare the simulated and observed original semimajor axes, $\sin 2\omega_G$ values, and previous perihelion passages of distant ($8<q<12$ au) LPCs. In Fig. 4A, we mark the parameter space within which a K-S test comparing simulations and observations cannot reject the null hypothesis (that the simulated and observed LPCs have the same underlying distribution) with at least 2$\sigma$ confidence.

Fig. 4A effectively marks the regions of LPC fading parameter space permitted by LPC observations and our simulations. The original semimajor axes, $\sin 2\omega_G$ values, and previous perihelia each define a region of permitted fading parameter space. We see that the LPC fading parameter spaces permitted by each of these LPC properties all broadly overlap with one another in Fig. 4A. In particular, we note that each parameter predicts that fading must begin beyond $\sim$12 au and that LPCs can make a maximum of $\sim$10 perihelion passages (and potentially as few as 2) before fading from observability. 

Previously, we found that our simulated $\sin 2\omega_G$ match to observed LPCs can be greatly improved if we truncate our simulated Oort cloud at $a=$ 25,000 au. Thus, in Fig. 4B, we study how the permitted LPC fading parameter space changes if we only consider LPCs from this truncated Oort cloud. Here we see that regions permitted by LPC original semimajor axes and previous perihelia shrink but still very much overlap. In particular, LPCs from this truncated Oort cloud must fade within $\sim$5 perihelion passages inside 20 au. We also note that the fading parameter space allowed by $\sin 2\omega_G$ expands dramatically. If simulated LPCs are only derived from Oort cloud semimajor axes between 20,000--25,000 au, this particular parameter puts no hard limit on the number of perihelion passages inside 20 au. However, the observed and simulated $\sin 2\omega_G$ values from our truncated Oort cloud can only agree if LPC fading initiates at least $\sim$12 au from the Sun.

\section*{Discussion}
The actual fading real LPCs experience is undoubtedly more complex than our assumed step function form, and the fading inferred in this work should also be taken to apply to LPCs of ``typical'' size. Very large or very small comet nuclei will presumably fade at differing rates owing to their differing supply of volatiles. For instance, Comet Hale-Bopp likely already defies our assumed fading law form, but its nucleus is also exceptionally large \cite{bail96,biv96,weav97}. Fading laws more complex than those explored here are fit to LPCs passing nearer to Earth \cite{weiss79,wietre99}, and it is very conceivable that additional distant LPC observations may require LPC fading different from ours. However, it is much less conceivable that observations of LPCs in the Saturn region can be explained without any kind of rapid fading whatsoever in the outer solar system.

LPC fading beyond 10 au undoubtedly complicates efforts to characterize cometary populations at large perihelia \cite{ivez19}. However, this fading may also provide new insight into LPCs. For instance, cometary activity is being discovered at extreme heliocentric distances, and the exact mechanism(s) and properties enabling it are not yet clear but may be constrained by the distance and nature of LPC fading \cite{womack2017supervolatiles,prialnik1990amorphous,meech2009activityobservations,hui2019}. 

In addition to generating distant ($8<q<12$ au) LPCs, our LPC production simulation generates near-Earth ($q<4$ au) LPCs as well, and we can also assess the distant fading incurred by near-Earth LPCs. To do this, we track the number of perihelion passages a ``new'' near-Earth ($q<4$ au, $a>10^4$ au) LPC makes within a given perihelion distance prior to its first near-Earth perihelion passage. Different near-Earth LPCs will of course make different numbers of passages within a given perihelion distance prior to inner solar system entry. In Figs. 4A and 4B, we plot the 80th percentile of the number passages made inside a given perihelion distance before our simulated near-Earth LPCs execute their first near-Earth perihelion passage. In Fig. 4A, we see that most near-Earth LPCs from our simulated Oort cloud avoid distant LPC fading prior to their arrival in the inner solar system, as their perihelion passage numbers mostly fall below the fading parameter space permitted by observations of distant LPCs. In contrast, in Fig. 4B we see that this is no longer the case if we truncate the Oort cloud's outer edge at $a=$ 25,000 au, as the perihelia of these lower semimajor axis LPCs evolve more slowly. Here we find that many near-Earth LPCs incur enough perihelion passages in the outer solar system to induce fading prior to their entry into the inner solar system. The differences between perihelion passage histories of near-Earth LPCs would likely become even more stark if our simulated Oort cloud extended interior to $a=$ 20,000 au. 

Thus, the distant LPC fading we document in this work may also have important ramifications for the near-Earth LPC population. If near-Earth LPCs are primarily derived from the Oort cloud's deep interior ($a\lesssim$ 20,000 au), we should expect most to make many perihelion passages 10--15 au from the Sun before evolving to a near-Earth ($q<4$ au) perihelion \cite{kaibquinn09}. Consequently, many of these ``new'' near-Earth ($q<4$ au, $a>10^4$ au) LPCs should have muted cometary activity at large heliocentric distances on their inbound leg into the inner solar system. On the other hand, near-Earth LPCs from the more distant parts of the Oort cloud will be less likely to undergo distant LPC fading before reaching a near-Earth perihelion and should display more vigorous cometary activity at large heliocentric distances. Observations assessing the fading status of inbound near-Earth LPCs beyond 5--10 au could help discern the ultimate source region of these bodies \cite{meech2009activityobservations}.

\section*{Methods and Materials}

\begin{table}
\begin{center}
\begin{tabular}{c c c c c c c}
Name & a$_{\rm orig}$ & q & i & $\Omega$ & $\omega$ & $\omega_{G}$ \\
 & (au) & (au) & ($^\circ$) & ($^\circ$) & ($^\circ$) & ($^\circ$) \\
 \hline
C/2000 A1 & 23068.07 & 9.74 & 24.55 & 111.89 & 14.34 & 33.34 \\
C/2001 G1 & 30127.20 & 8.24 & 45.36 & 203.94 & 343.33 & 65.69 \\
C/2003 A2 & 23468.66 & 11.42 & 8.07 & 154.57 & 346.60 & 47.34 \\
C/2003 S3 & 28790.03 & 8.13 & 151.52 & 226.28 & 154.38 & 191.65 \\
C/2004 T3 & 28633.38 & 8.87 & 71.92 & 50.39 & 259.68 & 220.54 \\
C/2007 D1 & 23303.53 & 8.79 & 41.43 & 171.13 & 340.12 & 43.51 \\
C/2007 K1 & 424.95 & 9.23 & 108.44 & 294.69 & 52.02 & 25.35 \\
C/2008 S3 & 50608.03 & 8.02 & 162.71 & 54.93 & 39.92 & 263.18 \\
C/2010 L3 & 5947.41 & 9.88 & 102.60 & 38.30 & 121.74 & 60.12 \\
C/2010 U3 & 16959.63 & 8.45 & 55.48 & 43.03 & 88.08 & 47.66 \\
C/2013 P3 & 24759.10 & 8.65 & 93.88 & 177.26 & 177.15 & 237.51 \\
C/2014 B1 & 16724.79 & 9.56 & 28.38 & 161.38 & 345.80 & 45.50 \\
C/2014 S1 & 72918.62 & 8.14 & 123.77 & 352.67 & 288.74 & 227.64 \\
C/2014 UN271 & 19565.16 & 10.95 & 95.47 & 190.00 & 326.28 & 25.49 \\
C/2015 D3 & 36785.36 & 8.14 & 128.55 & 156.94 & 2.72 & 80.23 \\
C/2016 C1 & 16880.44 & 8.46 & 56.20 & 181.78 & 328.58 & 33.91 \\
C/2016 E1 & 24742.63 & 8.16 & 132.07 & 233.05 & 47.48 & 79.53 \\
C/2017 AB5 & 15316.35 & 9.22 & 32.42 & 42.65 & 78.38 & 38.45 \\
C/2019 E3 & 34753.07 & 10.31 & 84.28 & 347.18 & 280.71 & 220.04 \\
C/2019 O3 & 26494.77 & 8.81 & 89.84 & 300.41 & 60.02 & 18.46 \\
C/2020 F2 & 16033.15 & 8.83 & 163.62 & 250.15 & 48.12 & 65.83 \\
C/2020 H5 & 7509.68 & 9.35 & 70.17 & 210.62 & 326.33 & 33.28 \\
C/2020 K2 & 7810.54 & 8.87 & 91.05 & 288.38 & 67.25 & 39.16 \\
C/2021 L3 & 9352.43 & 8.45 & 78.58 & 345.04 & 91.60 & 29.45 \\
C/2021 Q6 & 9940.67 & 8.70 & 161.84 & 133.45 & 140.83 & 267.47 \\
\hline
\end{tabular}
\caption{01-01-1950 osculating barycentric orbital elements of the 25 comets listed in the JPL Horizons system.  Columns are comet name designation, original semimajor axis, perihelion, ecliptic inclination, ecliptic longitude of ascending node, ecliptic argument of perihelion, and argument of perihelion relative to Milky Way plane.}
\end{center}
\end{table}

The JPL Horizons system currently lists 25 cometary objects with orbital periods over 200 years and perihelia between 8 and 12 au. These comets comprise our observational sample. Their ecliptic orbital elements (as well as Galactic arguments of perihelion) are listed in Table 1. To estimate the original semimajor axes of these long-period comets (LPCs), we used the JPL Horizons system to calculate their osculating barycentric orbits on 01-01-1950. Since these comets were all discovered after 2000, this places them well away from the planets and before they have incurred significant energy kicks on their current perihelion passage. For the subset of LPCs appearing in other catalogs of original semimajor axis, this yields general agreement \cite{mars08,krol17}.

Our simulation of long-period comet (LPC) production is performed with the SCATR integration package \cite{kaib11a}. $10^5$ massless test particles are integrated under the gravitational influence of the Sun, the four giant planets (on their modern orbits), and the Galactic tide for 2 Gyrs. Our formulation for the Galactic tide includes a radial term but is dominated by a vertical component whose strength is governed by the local density of the Milky Way disk, which we set to 0.1 M$_{\odot}$/pc$^3$ \cite{lev01,holmflynn00}. 

Initial particle eccentricities are drawn from a thermalized distribution uniform in $e^2$, but eccentricities are redrawn if the resulting initial particle perihelion is below 35 au. The reason for this redrawing of extreme eccentricities is that ejection via planetary perturbations leaves this portion of orbital space underpopulated relative to a pure thermal eccentricity distribution. We instead let the interplay of the Galactic tide and planetary perturbations slowly naturally populate the planetary region with LPCs as time progresses in our simulation. Initial particle inclinations are randomly selected from a uniform $\cos i$ distribution to reflect the isotropizing effects of stellar encounters and the Galactic tide \cite{dqt87}. Because the longitude of ascending node ($\Omega$) and argument of perihelion ($\omega$) cycling timescales under the Galactic tide for $a>$ 20,000 au are substantially shorter than the solar system age \cite{dqt87}, initial $\Omega$ and $\omega$ values (as well as mean anomalies) are randomly sampled from uniform distributions.

The initial semimajor axes of the test particles are assigned with a radial volume density profile proportional to $a^{-3.35}$ \cite{vok19} from a minimum semimajor axis of 20,000 au to a maximum semimajor axis of 50,000 au. The maximum semimajor axis is motivated by the very short dynamical lifetimes of Oort cloud bodies measured beyond this value \cite{kaib11b}. The minimum semimajor axis is less physically motivated. Although virtually all models of Oort cloud formation predict a substantial population of bodies interior to $a<20,000$ au \cite{dqt87,dones04,bras06,kaib11b}, the one major observational constraint on the Oort cloud's properties comes from the (predominantly near-Earth) LPC catalog. While an Oort cloud population interior to $a<20,000$ au can generate these observed LPCs, its presence is not strictly necessary to explain their existence \cite{kaibquinn09}. Meanwhile, our initial interest in distant LPC fading developed from the observed paucity of distant LPCs ($8<q<12$ au) with semimajor axes below $\sim$10$^4$ au. From prior work, it is clear that bodies in the inner 20,000 au of the Oort cloud will dramatically enhance the production of distant LPCs with $a\lesssim$ 10$^4$ au \cite{siltre16}. This means that inclusion of $a<20,000$ au orbits in our initial conditions would increase the likelihood that LPC fading is required to bring our simulations into agreement with observations. Thus, while unphysical, the choice to exclude $a<20,000$ au Oort cloud orbits is a conservative one with respect to assessing the need for LPC fading. 

The SCATR integration package is able to integrate particles in a barycentric frame using large time steps when they are far from the Sun and planets and then transition to a heliocentric integration with smaller time steps when particles approach the Sun and planets. In our simulation, we set this transition distance to $r=300$ au. In the heliocentric realm, we use a time step of 200 days, and in the barycentric realm, the step is increased to 3600 days. Particles are removed from the simulation if their heliocentric distance exceeds 1 pc or if they collide with the Sun or a planet.

Our simulation is integrated for 2 Gyrs, and every particle perihelion passage within 20 au of the Sun is recorded. For each perihelion passage, barycentric orbital elements are recorded when the particle is 50 au from the Sun on its inbound leg. These orbital element recordings form the basis of our simulated LPC distributions in the main paper. Each massless test particle is treated generically as a ``typical'' detectable LPC, and our simulations assume a 100\% detection efficiency (before comets fade). Of course, real LPCs have a range of nuclear radii as well as cometary activity levels, and their real detection efficiency is not 100\%. However, since all LPC orbits are nearly parabolic near perihelion, we do not expect observing biases to vary across our semimajor axis range of interest (the focus of Fig. 1), and there is no a priori reason to expect discovery probability to sharply spike near $\sin 2\omega_{G} = 1$ (the focus of Fig. 2). 

Moreover, the population of known, detected LPCs represents a subsample of LPCs that has a 100\% detection efficiency by definition. We seek to use our simulations to assess whether comet fading (detection efficiency collapsing) after limited perihelion passages is necessary to replicate the observed LPC population with $8<q<12$ au. In our work, we assume simulated LPCs transition from 100\% detection probability to 0\% detection probability after they fade. Although likely simplistic, LPC fading remains largely observationally uncharacterized, and this binary detectability approximation is used in other comet fading studies  \cite{wietre99, nes17, vok19}. What we demonstrate in our the main paper is that the known LPCs with $8<q<12$ au imply that a much larger companion population of LPCs should exist on similar perihelia but different semimajor axes, arguments of perihelion, and previous pre-discovery perihelion passages. Comet fading leading to their undetectability is the most plausible reason for their absence from the known LPC catalog.

Because our initial perihelion distribution is truncated at 35 au, the planet-crossing LPC population is unphysically low at the start of our simulation, and we do not consider LPCs that enter the inner 20 au of the solar system during the first 500 Myrs of our simulation. 500 Myrs is much longer than both the semimajor axis diffusion and perihelion torquing timescales of LPCs near $q\sim15$ au (which are both $\sim$10$^7$ years for our semimajor axes of interest \cite{dqt87}). Thus, the planetary system has a well-evolved set of LPC orbits when we begin recording them. Since the smallest semimajor axis of an observed LPC with $8<q<12$ au is 425 au, we only consider simulated LPCs with semimajor axes over 400 au. Our sample of simulated LPC orbits with $8<q<12$ au in the main paper is generated from the evolution of 1168 individual particles. Our sample of simulated near-Earth LPC orbits with $q<4$ au in the main paper is generated from the evolution of 594 individual particles.

\section*{References and Citations}
\bibliography{scibib}

\bibliographystyle{Science}

\section*{Acknowledgments:}
This work was greatly improved by the comments and suggestions from Julio A. Fern\'{a}ndez and one other anonymous reviewer. This work was performed with support from NSF CAREER Award 1846388 and a University of Oklahoma Junior Faculty Fellowship. Computing for this project was performed at the OU Supercomputing Center for Education \& Research (OSCER) at the University of Oklahoma (OU). NAK thanks the Case Western Reserve University Department of Astronomy for hosting him as a visitor during this work's preparation. NAK also thanks Rita Kaib for helpful discussions and a lifetime of companionship. Author contributions: NAK performed all work and analysis in this publication. Competing interests: The author declares that they have no competing interests. Data availability: All data needed to evaluate the conclusions in the paper are present in the paper and/or the Supplementary Materials. The simulation dataset and analysis codes necessary to replicate this work are available on the publicly accessible repository, figshare, at https://doi.org/10.6084/m9.figshare.19125413. 

\section*{List of Supplementary Materials:}
Supplementary Text\\
Figs. S1\\
References \textit{(32--37)}

\clearpage

\setcounter{page}{1}

\section*{Supplementary Text}
\renewcommand{\thefigure}{S\arabic{figure}}
\setcounter{figure}{0}  

\subsection*{Non-Gravitational Forces}

Non-gravitational forces arising from jetting in cometary activity are known to alter the orbits of comets passing near the Sun \cite{mars73}. However, these forces are not included in our simulations or accounted for in our analysis of observed LPCs with $8<q<12$ au. These forces have been suspected to be relevant even for distant comets  \cite{jew21}, and the rapid fading of distant LPCs documented in the main paper supports the idea that non-gravitational forces at large heliocentric distances may be stronger than previously suspected. Among currently known distant LPCs, these forces' characterization is uncertain \cite{jew21}. The absence of distant LPCs with hyperbolic original orbital solutions is consistent with smaller non-gravitational forces among distant LPCs compared to LPCs nearer to Earth. On the other hand, the known distant LPCs exist at the edge of our detection capabilities and are thus likely to have larger than average nuclei, minimizing the effects of non-gravitational forces. Regardless, because of the dramatic deviations between observed and simulated LPC orbits in the main paper, the addition of non-gravitational forces is very unlikely to obviate the need for LPC fading in the outer solar system. 

\subsection*{Stellar Passages}

In the simulation presented in the main paper, we made the decision to omit stellar passages. The reason for this is that our simulated LPCs are compiled by analyzing the flux of bodies through the planetary region over the course of 1.5 Gyrs. Meanwhile, with a median semimajor axis of $\sim$23,000 au, the observed distant LPCs really should only contain information about the past few Myrs of stellar encounters \cite{rick12}. Although we could monitor the near-Earth LPC flux in our simulations to omit obvious comet shower periods wherein the Oort cloud was perturbed by rare ($<$ once per 100 Myrs) powerful stellar passages, the period that our simulated LPC statistics spanned would still contain a much greater number of more intermediate, but still impactful passages that are unlikely to have affected the recent observed LPC flux. 

However, it may still be instructive to assess the potential effects of stellar encounters on our results, so in this supplementary section, we repeat our main paper's simulation, but we now include a population of passing field stars. In this second simulation, the 1 pc surrounding the solar system is penetrated by $\sim$18 field star encounters per Myr \cite{bail18}. For these encounters, stellar masses are assigned via the locally observed present day mass function \cite{reid02} and incoming velocities, which we assume to be isotropic, are assigned in a mass-dependent manner derived from the observed kinematics of the solar neighborhood \cite{rick08,gar01}. 

\begin{figure}
\centering
\includegraphics[scale=0.4]{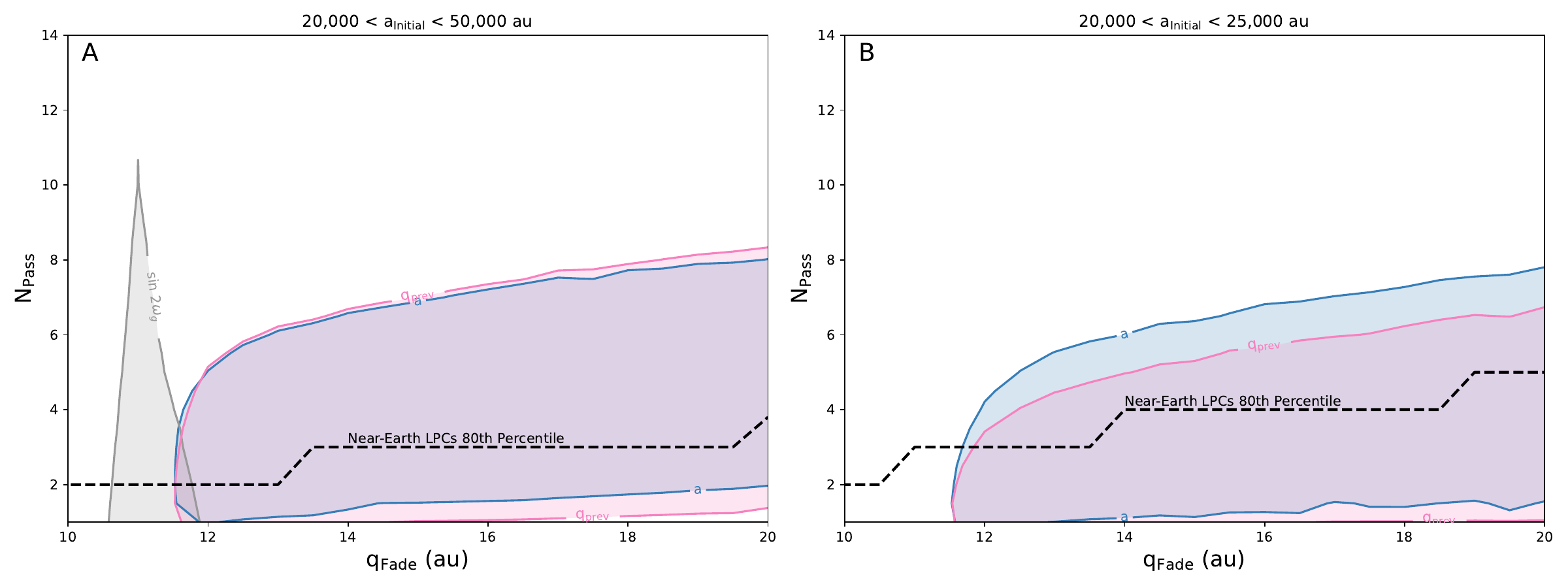}
\caption{{\bf Distant long-period comet fading parameter space permitted by known comets and simulated comet production that includes field star passages.} Number of perihelion passages before comets fade are plotted against the perihelion distance inside which fading operates for LPCs with $8<q<12$ au. Filled contours mark area outside of which a K-S test rejects (with over 2$\sigma$ confidence) the null hypothesis that simulated and observed LPCs have the same underlying distribution. Contours are calculated for LPC original semimajor axis, $\sin 2\omega_G$, and estimated previous perihelion passage. For simulated new ($a>10^4$ au) near-Earth ($q<4$ au) LPCs, we also plot the 80th percentile of the number of passages made inside a given perihelion prior to inner solar system entry ({\it dashed line}). Panel {\bf A} considers all simulated comets and Panel {\bf B} excludes those initialized on semimajor axes beyond 25,000 au.}
\end{figure}

With this new simulation, we now replicate the analysis of Fig. 4 of the main paper, which is shown in Fig. S1. Here we see that boundaries of fading parameter space allowed by our simulated $\sin 2\omega_{G}$ distribution change dramatically, in that there are almost no regions that provide an adequate fit to observations. This is not surprising, since, as we explain above, the LPCs in this simulation are likely less influenced by the Galactic tide than the actual solar system's LPCs. Thus, no matter how extreme we make LPC fading, we cannot generate as strong of a skew toward $\sin 2\omega_{G} = 1$ as the observed LPCs display. 

In contrast, the area of fading parameter space allowed by LPC semimajor axis and previous perihelion passage distance is not as impacted as strongly. Comparing the main paper's Fig. 4A with this new figure, we see that our new simulation has a lower maximum number of passages allowed before LPCs fade. This is also expected, since the inclusion of stellar perturbations increases the rate of LPC perihelion change. To prevent a larger number of small semimajor axis LPCs from being observed, fading must be somewhat stronger beyond the 12 au limit of our observed sample.

\end{document}